\documentstyle[12pt]{article}
\def\be{\begin{eqnarray}}
\def\ee{\end{eqnarray}}
\def\ba{\begin{array}}
\def\ea{\end{array}}
\def\G{{\cal G}}
\def\B{{\cal B}}
\def\A{{\cal A}}
\def\M{{\cal M}}
\def\T{{\cal T}}
\def\L{{\cal L}}
\def\X{{\cal X}}
\def\Z{{\cal Z}}

\def\C{{\cal C}}

\def\J{{\cal J}}
\def\E{{\cal E}}
\def\F{{\cal F}}
\begin{document}
\begin{center}
{\LARGE{New progress in the stationary \\
\vskip 5mm
D=N=4 supergravity}}
\end{center}
\vskip 25mm
\begin{center}
{\bf \large {Oleg V. Kechkin}}
\end{center}
\begin{center}
DEPNI, Institute of Nuclear Physics\\
M.V. Lomonosov Moscow State University\\
119899 Moscow, Vorob'yovy Gory, Russia\\
e-mail:\, kechkin@depni.npi.msu.su
\end{center}
\vskip 25mm
\begin{abstract}
A bosonic sector of the four-dimensional low-energy heterotic string
theory with two Abelian gauge fields is considered in the stationary
case. A new $4\times 4$ unitary null-curvature matrix representation
of the theory is derived and the corresponding formulation based on
the use of a new $2\times 2$ Ernst type matrix complex potential is
developed. The group of hidden symmetries is described and classified
in the matrix-valued quasi General Relativity form. A subgroup of
charging symmetries is constructed and representation which transforms
linearly under the action of this symmetry subgroup is established.
Also the solution generation procedure based on the application of the
total charging symmetry subgroup to the stationary Einstein-Maxwell
theory is analyzed.
\end{abstract}
\vskip 10mm
\begin{center}
\end{center}
\newpage
\section{Introduction}

String theories provide us new physically interesting examples of the
field theories which describe some concrete sets of matter fields
coupled
to gravity \cite{Kir}. In this paper we consider a bosonic sector of
the effective field theory arising in the low-energy limit of heterotic
string theory. In \cite{MahSch} it was performed a general toroidal
compactification of this theory; in \cite{Sen3} it was shown that in
the case of compactification to three dimensions the theory becomes a
symmetric space model coupled to gravity on shell. This symmetric space
model parameterizes the coset $O(d+1,d+1+n)/O(d+1)\times O(d+1+n)$,
where $d$ and $n$ are the numbers of compactified space-time
dimensions and original Abelian gauge fields respectively. The
corresponding non-linear $\sigma$-model allows different null-curvature
matrix representations \cite{Sen3}, \cite{HKMEP}. One of them
(\cite{HKMEP}) is closely related to the formulation of theory in
terms of pair of the real matrix potentials which are the
$(d+1)\times (d+1)$ and $(d+1)\times (d+1+n)$ matrix fields. The matrix
potential formulation generalizes the conventional complex potential
one of the stationary Einstein-Maxwell theory \cite{ErnstEM} to the
heterotic string theory case. This formulation is especially useful for
the study of hidden symmetries of the theory; in \cite{HKCS} it was
explored for classification of the general low-energy heterotic string
theory
in the quasi Einstein-Maxwell form and for the determination of the
three-dimensional charging symmetry transformations. In \cite{ZF} it was
developed in details a new formalism based on the use of the new
$(d+1)\times (d+1+n)$ real matrix potential which undergoes linear
transformations when the charging symmetry subgroup acts. This subgroup
leaves the trivial values of the all three-dimensional fields unchanged
and forms a base for any procedure of the symmetry generation of the
asymptotically flat solutions. The most general generation technique of
this type is developed in \cite{ZF} and, in fact, the problem of the
most general symmetry extension of the three-dimensional asymptotically
flat solutions in heterotic string theory is solved in this work.

However, these general results possess a qualitative modernization in
the following sense. For some special values of the parameters $d$ and
$n$ one can develop more compact alternative formulations of the
theory than the general orthogonal ones which exists for the arbitrary
$(d,n)$-labeled  representative of the heterotic string series of the
theories. For example, in the case of $d=n=1$ \, (the Einstein-Maxwell
dilaton-axion theory, see \cite{GKSP}, \cite{GKMEP}) it is possible to
use the compact $4\times 4$ symplectic null-curvature matrix
representation instead of the $5\times 5$ orthogonal one. Then, if
$d=2, n=0$ (the five-dimensional dilaton-Kalb-Ramond gravity), the
same situation takes place, but in this case it arises the
$SL(4,R)/SO(4)$ matrix representation. In this paper we show, that
for the theory with $d=1,n=2$,
in addition to the conventional orthogonal $6\times 6$ null-curvature
matrix, one can use the $4\times 4$ one which parameterizes the coset
$SU(2,2)/S[U(2)\times U(2)]$. We derive the parameterization of this
matrix in terms of components of the heterotic string theory
fields and relate the new null-curvature matrix to the $2\times 2$ complex
matrix potential of the Ernst type. Thus, we show that this theory,
which also arises in framework of the $D=N=4$ supergravity
(see \cite{KalIWP}, \cite{Sab} for the extremal and
black hole solutions), possesses
the formal properties of the stationary General Relativity. In
particular, we 	separate the group of the three-dimensional hidden
symmetries to the matrix-valued shift, scaling and Ehlers parts
(see \cite{EMT} for comparison with the General Relativity analogies).
We also establish another $2\times 2$ complex matrix potential which
transforms linearly in respect to the general group of charging
symmetry transformations. These new representations derived
in this paper for the $d=1,n=2$ theory maximally simplify and
compactify the results developed for the general low-energy heterotic
string theory systems. The new unitary null-curvature matrix based
description of the theory seems especially promising for
application of the inverse scattering transform method for construction
of the soliton solutions \cite{BZ}, \cite{EGK}.

\section{A symmetric space model}

In the string theory frame the action of the low-energy heterotic
string theory reads \cite{Kir}:
\be\label{s1}
S_4=\int d^4X |{\rm det}G_{MN}|^{\frac{1}{2}}\,e^{-\Phi}
\left ( R_4+\Phi_{,M}\Phi^{,M}-\frac{1}{12}H_{MNK}H^{MNK}-
\frac{1}{4}F^I_{MN}F^{I\,MN}\right ),
\ee
where $H_{MNK}=\partial_MB_{NK}-\frac{1}{2}A^I_MF^I_{NK}+\,{\rm cyclic}\,
\{M,N,K\}$ and $F^I_{MN}=\partial_MA^I_N-\partial_NA^I_M$. Here
$M,N,K=0,...,3$, the space-time signature is $(-,+,+,+)$ and $I=1,2$.
In the stationary case all the fields are $t$-independent, where
$t=X^0$; let us classify the field components in respect to the
transformations of the three-dimensional coordinates
$x^{\mu}=X^{\mu}$. There is the set of the scalars:
\be\label{01}
G=G_{00},\,\,\, A=(A_0^1\,\,A_0^2),\,\,\, 
\phi=\Phi-\log{|G|^{\frac{1}{2}}}.
\ee
Then, there are three vectors with the components
\be\label{02}
V_{1\mu}=G^{-1}G_{0\mu}, \, \,
V_{2\mu}=B_{0\mu}+\frac{1}{2}A_0^IV_{3\mu}^I \ ,\,
V_{3\mu}^I=-A^I_{\mu}+A_0^IV_{1\mu}.
\ee
Finally, there are two tensor
fields
\be\label{03}
h_{\mu\nu}=e^{-2\phi}(G_{\mu\nu}-GV_{1\mu}V_{1\nu}),\,\,\,\,\, 
b_{\mu\nu}=B_{\mu\nu}-\frac{1}{2}(V_{1\mu}V_{2\nu}-V_{1\nu}V_{2\mu}).
\ee
In fact the field $b_{\mu\nu}$ is non-dynamical; following \cite{Sen3}
we put it equal to zero without imposing of any additional restrictions
on the other degrees of freedom. The following use of the motion
equations allows one to introduce three pseudo scalar fields $u,v$ and $s$
\,\, ($s^T=(s^1\,\,s^2)$) accordingly the relations
\be\label{04}
&&\nabla\times\vec
V_1=e^{2\phi}(\nabla u+\frac{1}{2}AA^T\nabla v+A\nabla s), 
\nonumber\\
&&\nabla\times\vec V_2=e^{2\phi}G\nabla v-\frac{1}{2}AA^T
\nabla\times\vec
V_1+A\nabla\times\vec V_3, 
\nonumber\\
&&\nabla\times\vec V_3=e^{2\phi}(\nabla
s+A^T\nabla v)+A^T\nabla\times\vec V_1.
\ee
The resulting effective
three-dimensional theory can be expressed in terms of the three $2\times
2$ matrices $\G$, $\B$ and $\A$
of the following form \cite{HKMEP}:
\be \label{s3}
\G=
\left(
\ba{cc}
-e^{-2\phi}+Gv^2&Gv\cr
Gv&G
\ea
\right),
\quad
\B=
\left(
\ba{cc}
0&-(u+\frac{1}{2}As)^T\cr
u+\frac{1}{2}As&0
\ea
\right),
\quad
\A=\left(
\ba{c}
s^T+vA \cr
A
\ea
\right).
\ee
The corresponding three-action reads:
\be \label{s4}
S_3=\int d^3x h^{\frac{1}{2}}\left ( -R_3+L_3\right ),
\ee
where
\be\label{s5}
L_3={\rm Tr}\left [ \frac{1}{4}\left (\J_{\G}^2-\J_{\B}^2\right )
+\frac{1}{2}\nabla\A\nabla\A^T\G^{-1}\right ]
\ee
and $\J_{\G}=\nabla\G\,\G^{-1}, \,\, \J_{\B}=\nabla\B\,\G^{-1}$. To
clarify the symmetric properties of the theory it is useful to introduce
the following $6\times 6$ real matrix $\M$:
\be \label{s6}
\M=
\left(
\ba{ccc}
\G^{-1}&\G^{-1}(\B+\T)&\G^{-1}\A\cr
(-\B+\T)\G^{-1}&(\G-\B+\T)\G^{-1}(\G+\B+\T)&(\G-\B+\T)\G^{-1}\A\cr
\A^T\G^{-1}&\A^T\G^{-1}(\G+\B+\T)&1+\A^T\G^{-1}\A
\ea
\right),
\ee
where $\T=\frac{1}{2}\A\A^T$. This matrix satisfies the coset relations
\be \label{s7}
\M^T=\M, \qquad \M\L\M=\M,
\ee
where
\be \label{s8}
\L=
\left (
\ba{ccc}
0&1&0\cr
1&0&0\cr
0&0&-1
\ea
\right).
\ee
The restrictions (\ref{s7}) are preserved by the transformation
\be\label{s9}
\M\rightarrow\C^T\M\C,
\ee
where
\be\label{s10}
\C^T\L\C=\L.
\ee
The Lagrangian $L_3$, being expressed in terms of $\M$, reads:
\be\label{s11}
L_3=\frac{1}{8}{\rm Tr}\left (\J_{\M}\right )^2,
\ee
where $\J_{\M}=\nabla\M\,\M^{-1}$; it is invariant under the action of
the
transformation (\ref{s9})-(\ref{s10}). Thus, this transformation is a
symmetry of the theory. In \cite{Sen3} it was originally shown that this
map gives a total group of the three-dimensional symmetry
transformations.

Our approach is oriented to the close analogy between the heterotic
string and Einstein-Maxwell theories. For example, this analogy
``generates'' the fact that the Lagrangian (\ref{s5}) takes the
conventional Einstein-Maxwell form if $\G=f, \,\, \B=\chi\epsilon, \,\,
\A=V+U\epsilon$, where $\epsilon$ is the antisymmetric $2\times 2$
matrix ($\epsilon_{12}=1$) and the quantities $f, \chi, U$ and $V$ are
the (nonmatrix) functions. Actually, for the above given choice the
motion equations of the theory can be derived from the following
effective three-dimensional Lagrangian:
\be\label{s12}
L_3=\L_{EM}=\frac{1}{2f^2}\left |\nabla \E-\bar \F\nabla \F\right |^2-
\frac{1}{2}\left |\nabla \F\right |^2,
\ee
where $\E=f+i\chi-\frac{1}{2}(V^2+U^2), \,\, \F=V+iU$. It is exactly the
Lagrangian of the stationary Einstein-Maxwell theory; the complex
functions $\E$ and $\F$ play the role of the conventional Ernst potentials
\cite{ErnstEM}. Below we show that, moreover, the complete theory under
consideration possesses a representation in the pure Einstein theory
form. Namely, we derive a formulation of this theory in terms of the
single complex $2\times 2$ matrix potential of the Ernst type. This fact
leads to the crucial formalism compactification and opens
new possibilities in
concrete applications of the different powerful methods of solution
construction.
Actually, from the general theory point of
view there are two real $2\times 2$ matrix Ernst potentials $\X$ and
$\A$, where $\X=\G+\B+\frac{1}{2}\A\A^T$ \cite{HKCS} which alternatively
define the $6\times 6$ orthogonal symmetric space matrix $\M$
(see (\ref{s6})).
However, as it is shown below, it exists the single $2\times 2$ complex
Ernst matrix potential $E$ which is uniquely related to some new
unitary $4\times 4$ symmetric space matrix $M$. This new unitary
null-curvature matrix representation is actually convenient in framework
of application of the inverse scattering transform method \cite{BZ} to
the theory under consideration after its following reduction to two
dimensions.


\section{New Ernst potential formulation}

A general classification of the symmetric space models coupled to
gravity in three dimensions is given in \cite{BGM}. Let us consider
the
one with the three-dimensional Lagrangian
\be\label{e1}
L_c=\frac{1}{4}\left ( J_M\right )^2,
\ee
where $J_M=\nabla M\,\, M^{-1}$. For us it will be
important to restrict the $4\times 4$ null-curvature matrix $M$
by the following unitary coset space relations:
\be\label{e2}
M^+=M,\quad M LM,
\ee
where
\be\label{e2'}
L=
\left (
\ba{cc}
0&1\cr
1&0
\ea
\right).
\ee
The action (\ref{e1}) is invariant in respect to the transformation
$M\rightarrow  C^+M C$, where $ C^+ L
 C= L$, so we deal with the symmetric space model with the
unitary hidden symmetry. Our statement is the following: this new
unitary $\sigma$-model coincides with the orthogonal one introduced in
the previous section. This means that the matrix $M$ can be
parameterized by the quantities $G,A,\phi,u,v,$ and $s$ in such a way
that  this parameterization will give the general solution of the coset
restrictions (\ref{e2}) and also it will guarantee the equality
$L_c=L_3$. The rest part of this section is related to determination of
the explicit form of this new unitary representation of the theory.

First of all, let us solve the unitary coset restrictions (\ref{e2});
the result reads:
\be\label{e3}
M=
\left (
\ba{cc}
 P^{-1}&i P^{-1}Q\cr
-iQ P^{-1}& P+Q P^{-1}Q
\ea
\right),
\ee
where $ P^+= P$ and $Q^+=Q$ are the $2\times 2$
complex matrices. A substitution of Eq. (\ref{e3}) to Eq. (\ref{e1})
shows that
\be\label{e4}
L_c=\frac{1}{2}\left ( \J_ P^2+\J_Q^2\right ),
\ee
where $\J_ P=\nabla P\,\, P^{-1}$ and
$\J_Q=\nabla Q\,\, P^{-1}$. We plan to calculate the
traces in Eqs. (\ref{s5}) and (\ref{e4}) and to comparize the
Lagrangians $L_3$ and $L_c$. For our purposes it will be convenient
to explore the following (temporary) parameterizations of the matrices
$\G,\B,\A$ and $ P,\,Q$:
\be\label{e5}
&&\G=-g_0
\left (
\ba{cc}
g_1^{-1}+g_1g_2^2&g_1g_2\cr
g_1g_2&g_1
\ea
\right),\,\,\,
\B=
\left (
\ba{cc}
0&-b\cr
b&0
\ea
\right),\,\,\,
\A=
\left (
\ba{cc}
a_1&a_3\cr
a_4&a_2
\ea
\right),\nonumber\\
&& P=p_0
\left (
\ba{cc}
-p_1^{-1}+p_1|p_2|^2&p_1\bar p_2\cr
p_1p_2&p_1
\ea
\right),\,\,\,\,\,\,
 P=p_0
\left (
\ba{cc}
q_0&q_2\cr
\bar q_2&q_1
\ea
\right).
\ee
Here the parameters $p_2=p_2^{'}+ip_2^{''}$ and $q_2=q_2^{'}+iq_2^{''}$
are complex; all the other ones are real. The
parameterization (\ref{e5}) essentially simplifies the comparison
procedure. Note, that the signature in $ P$ is taken indefinite
because the equality $L_3=L_c$ becomes possible only in this case. After
some algebra for the comparing Lagrangians one obtains:
\be\label{e6}
&&L_3=\frac{1}{2}\left\{
g_0^{-2}\nabla g_0^2+g_1^{-2}\nabla g_1^2+g_1^2\nabla g_2^2+g_0^{-2}
\left [
\nabla b+\frac{1}{2}\left ( a_4\nabla a_1-a_1\nabla a_4+
a_2\nabla a_3-a_3\times\right.\right.\right.
\nonumber\\
&&\times\left.\left.\left.\nabla a_2\right )\right ]^2-g_0^{-1}g_1\left [
\left (\nabla a_1-g_2\nabla a_4\right )^2
+\left (\nabla a_3-g_2\nabla a_2\right )^2\right ]-g_0^{-1}g_1^{-1}
\left [ \nabla a_2^2+\nabla a_4^2\right ]\right\},
\nonumber\\
&&L_c=
p_0^{-2}\nabla p_0^2+p_1^{-2}\nabla p_1^2+\frac{1}{2}\left\{
p_0^{-2}p_1^{-2}\nabla q_1^2+p_0^{-2}p_1^2\left [
\nabla q_0-2\left ( p_2^{'}\nabla q_2^{'}-p_2^{''}\nabla q_2^{''}\right
)+\left ( p_2^{'\,2}+\right.\right.\right.
\nonumber\\
&&\left.\left.\left.
+p_2^{''\,2}\right )\nabla q_1
\right ]^2\right\}
-p_0^{-2}\left [ \left (\nabla q_2^{'}-p_2^{'}\nabla q_1
\right )^2+
\left (\nabla q_2^{''}+p_2^{''}\nabla q_1\right )^2\right ]-p_1^2
\left [ \nabla p_2^{'\,2}+\nabla p_2^{''\,2}\right ].
\ee
Now the identification process can be performed in the following way.
From comparison of the coefficients before the brackets in the last
four `negative' terms one concludes that $g_0^{-1}g_1^{-1}\sim p_1^2$
and $g_0^{-1}g_1\sim p_0^{-2}$. In this case two first `positive' terms
are equal with the arbitrary choice of the coefficients of
proportionality. Let us put
\be\label{e7}
g_0=p_0p_1^{-1}, \,\, g_1=p_0^{-1}p_1^{-1}, \,\, g_2=q_1;
\ee
then three first terms coincide evidently. It is easy to see that if
\be\label{e8}
&&a_1=\sqrt 2\left (q_2^{'}-p_2^{'}q_1\right ),\,\,
a_2=\sqrt 2p_2^{''},\nonumber\\
&&a_3=\sqrt 2\left (q_2^{''}+p_2^{''}q_1\right ),\,\,
a_4=-\sqrt 2p_2^{'},
\ee
then the four last terms also become equal. The last step is to
substitute Eqs. (\ref{e7}) and (\ref{e8}) to fourth term of $L_3$ in Eq.
(\ref{e6}). This allows one to compare the remaining non-identified
parameters $b$ and $q_0$. The result reads:
\be\label{e9}
b=q_0-p_2^{'}q_2^{'}+p_2^{''}q_2^{''}.
\ee
Thus, Eqs. (\ref{e7})-(\ref{e9}) give the solution of the our problem.
Using them and Eqs. (\ref{s3}) and (\ref{e5}) one finally has:
\be\label{e10}
P\!=\!
\frac{e^{-\phi}}{|G|^{\frac{1}{2}}}\!
\left (
\ba{cc}
\!-\!G\!-\!\frac{1}{2}[(A_0^1)^2\!+\!(A_0^2)^2]&\frac{1}{\sqrt 2}
(A_0^1\!+\!iA_0^2)\cr
\frac{1}{\sqrt 2}(A_0^1\!-\!iA_0^2)&-1
\ea
\right), \,\,
Q\!=\!
\left (
\ba{cc}
u&\frac{1}{\sqrt 2}(s^1\!+\!is^2)\cr
\frac{1}{\sqrt 2}(s^1\!-\!is^2)&v
\ea
\right).
\ee
It is interesting to note, that the matrix $ P$ is constructed
from the scalar field components, whereas $Q$ is defined by the
pseudo scalar ones only. This opportunity essentially simplifies the
translation of the solution obtained in the $\sigma$-model terms to
the language of the heterotic string theory field components.


\section{Study of hidden symmetries}

Now let us introduce the potential
\be\label{g1}
E= P+iQ;
\ee
then the effective three-dimensional matter Lagrangian $L_c\equiv
L_3$ takes the following form:
\be\label{g2}
L_3=2{\rm Tr}\left [\nabla E\left (E+E^+\right )^{-1}
\nabla E^+\left (E+E^+\right )^{-1}\right ].
\ee
It is easy to see that the scale transformation
\be\label{g3}
E\rightarrow  S^+E S
\ee
is a symmetry for the arbitrary non-degenerated constant matrix
$ S$. Then, for the arbitrary Hermittean constant matrix $ R$
the shift
\be\label{g4}
E\rightarrow E+i R
\ee
is also a symmetry. To establish the rest part of the symmetry
transformations let us note that the map
\be\label{g5}
E\rightarrow E^{-1}
\ee
is a discrete symmetry of $L_3$. This map transforms the scale
transformation (\ref{g3}) to itself (with the re-parameterization
$ S\rightarrow ( S^{+})^{-1}$), whereas from the shift
transformation (\ref{g4}) one obtains the following new symmetry:
\be\label{g6}
E^{-1}\rightarrow E^{-1}+i L,
\ee
where $ L^+= L$. From Eqs. (\ref{g4}) and (\ref{g6}) it
follows that the map (\ref{g5}) relates the shift and this new nonlinear
symmetry by
the replacement $ R\leftrightarrow  L$.

Eqs. (\ref{g3}), (\ref{g4}) and (\ref{g6}) give the total group of the
hidden symmetry transformations of the theory under consideration. In
fact this group coincides with the unitary group discussed in the
previous section \, (in Eq. (\ref{g3}) the common phase multiplier in $ S$
is not sufficient). The transformations (\ref{g3}), (\ref{g5}) and
(\ref{g5}) are the straightforward matrix generalizations of the
scale, shift and Ehlers symmetries of the stationary General Relativity
(see \cite{Ehl}, \cite{Kin} and \cite{EMT}). In this remarkable analogy
the quantity $E$ plays the role of the matrix Ernst potential.

From Eqs. (\ref{e10}) and (\ref{g1}) it follows that the trivial value
of the all scalar and pseudo scalar three-fields leads to the special
value $E_0=\sigma_3$ of the matrix Ernst potential $E$. Let
us establish the explicit form of subgroup of the three-dimensional
charging symmetries, i.e., the form of the symmetry transformations which
preserve the above defined three-dimensional trivial field
configuration. To do this, let us introduce the following new matrix
potential $ Z$:
\be\label{g7}
 Z=2\left ( E+\sigma_3\right )^{-1}-\sigma_3.
\ee
Then after some algebra from Eq. (\ref{g2}) one obtains that
\be\label{g8}
L_3=2{\rm Tr}\left [ \nabla  Z\left (\sigma_3- Z^+\sigma_3
 Z\right )^{-1}
\nabla  Z^+\left
(\sigma_3- Z\sigma_3 Z^+\right )^{-1}
\right ].
\ee
It is easy to see that the transformation
\be\label{g9}
 Z\rightarrow e^{i\alpha} C_L Z C_R
\ee
describes a symmetry of Eq. (\ref{g8}) if the parameter $\alpha$ is real
and both the $ C_L$ and $ C_R$ matrices satisfy the $SU(1,1)$
group restrictions:
\be\label{g9'}
 C_L^+\sigma_3 C_L= C_R^+\sigma_3 C_R=\sigma_3,\quad
{\rm det}\, C_L={\rm det}\, C_R=1.
\ee
From comparison with the orthogonal group realization of the charging
symmetry subgroup of transformations given in \cite{HKCS} it immediately
follows that Eqs. (\ref{g9}) and (\ref{g9'}) give the general
charging symmetry subgroup of the theory.

Now let us briefly discuss the possible application of the established
unitary formulation to the problem of generation of the asymptotically
flat solutions. Such a generation is based on the use of
charging symmetries, because the asymptotically flat solutions are
trivial at the spatial infinity and only the subgroup of charging
symmetry transformations preserve this property.  It is easy to see that
for the trivial solution $ Z= Z_0=0$, and this special value is
actually preserved by the linear and homogenios transformation
(\ref{g9}). It is possible to start our generation from the subsystem
with
\be\label{g10}
 Z=\left (
\ba{c}
z\cr
0\cr
\ea
\right),
\ee
where $z=(z_1\,\,z_2)$. It is easy to verify that Eq. (\ref{g10})
actually defines a subsystem, i.e. that the anzats under consideration
is actually consistent. One can prove that the corresponding motion
equations can be obtained from the effective Lagrangian
\be\label{g11}
L_{EM}=2\,\frac{\nabla z \left (\sigma_3-z^+z\right )^{-1}\nabla z^+}
{1-z\sigma_3z^+},
\ee
which exactly coincides with the one given by Eq. (\ref{s12}). Actually,
the substitution (see also \cite{Maz})
\be\label{g11'}
z_1=\frac{1-E}{1+E}, \qquad z_2=\frac{\sqrt 2F}{1+E}
\ee
gives a proof of this statement immediately. Thus, the subsystem
(\ref{g10}) coincides with the stationary Einstein-Maxwell theory and
one can start from the asymptotically flat Einstein-Maxwell fields for
generation of the heterotic string theory solutions accordingly Eq.
(\ref{g9}). In this generation one can omit the transformation subgroup
given by the phase factor $e^{i\alpha}$ and the group matrix $ C_R$
if one starts
from the charging symmetry complete Einstein-Maxwell fields. Actually,
it is easy to see that the map
\be\label{g11''}
z\rightarrow e^{i\alpha}z C_R
\ee
is a symmetry of Eq. (\ref{g11}) which preserve the trivial $z$-value.
In fact it is exactly the subgroup $U(1)\times SU(1,1)$ of the
Einstein-Maxwell theory charging symmetries \cite{EMT}. Thus, in this
sense, the most general procedure of the continuous symmetry extension
of the
asymptotically flat Einstein-Maxwell solutions to the field of heterotic
string theory under consideration is given by Eqs. (\ref{g9}) and
(\ref{g10}) with arbitrary $ C_L\in SU(1,1)$ and $ C_R=1$,
$\alpha=0$. We hope to present explicit generation examples in the
forthcoming publications. At the end of this section let us note that
the formulation  of the theory based on the use of the complex matrix
potential $ Z$ allows one to obtain results in the charging
symmetry invariant form. This statement is related to the both symmetry
technique application and straightforward solution construction. In the
latter case one must use an anzats which is given by restrictions
invariant in respect to the charging symmetry transformation (\ref{g9}).


\section*{Acknowledgments}
This work was supported by RFBR grant ${\rm N^{0}}
\,\, 00\,02\,17135$.


\section{Conclusion}

The main results of this paper are the $E$- and $ Z$-based
matrix potential formalisms of the theory. We have found an explicit
form of the parameterization of the Ernst matrix potential $E$
in terms of the heterotic string theory physical field components. Also
we have studied the group of hidden symmetries of the theory in
framework of these new approaches. In this study the matrix potential
$ Z$ is especially convenient for the work with asymptotically
flat solutions. We have also indicated a possibility of generation of
the heterotic string theory solutions starting from the stationary
Einstein-Maxwell fields. The results obtained in this paper concludes
the investigation performed in \cite{OKU}. The $\Z$-formalism developed
in this paper is explicitly related to the unitary null-curvature matrix
formulation and seems really promising for the following application of
the inverse scattering transform method to this theory. This statement
is more than only a hope in view of the role which the conventional
Ernst potential formulation play in construction of the soliton solution in
the classical General Relativity. Now the use of the matrix Ernst potential
$E$ guarantees the positive result for this activity in the field
of the discussed four-dimensional heterotic string theory. This result
is controlled by the close analogy between the Einstein and heterotic
string theories in three and lower space-time dimensions. The
generalization of the obtained results to the case of the heterotic
string theory with the arbitrary numbers of toroidally compactified
dimensions and Abelian gauge fields can be performed in framework of the
special projective formalism developed in \cite{OKPF}. This
generalization will also possess the charging symmetry completeness
property for the generated asymptotically flat solutions.


\end{document}